\documentclass[11pt]{article}
\usepackage{mathrsfs}
\usepackage{amsmath}
\usepackage{amsfonts}
\usepackage{amssymb, amsmath, cite}
\usepackage{color}
\usepackage{graphicx}

\newcommand\be{\begin{equation}}
\newcommand\ee{\end{equation}}
\newcommand\ber{\begin{eqnarray}}
\newcommand\eer{\end{eqnarray}}
\newcommand\berr{\begin{eqnarray*}}
\newcommand\eerr{\end{eqnarray*}}
\newcommand\bea{\begin{eqnarray}}
\newcommand\eea{\end{eqnarray}}

\newcommand\dd{\mbox{d}}

\newcommand{\dt}{\dot{t}}
\newcommand{\nn}{\nonumber}
\newcommand{\dr}{\dot{r}}

\title{Determination of Bending Angle of Light Deflection Subject to Possible Weak and Strong Quantum Gravity Effects}

\author{Chenmei Xu\\School of Mathematics and Statistics\\Henan University\\
Kaifeng, Henan 475004, PR China\\ \\
Yisong Yang\footnote{Corresponding author. Email address: yisongyang@nyu.edu}\\Courant Institute of Mathematical Sciences\\ New York University\\New York, New York 10012, USA
}

\date{}

\begin{document}
\maketitle
\begin{abstract}
Explicit expressions for the bending angle of light deflection arising from phenomenologically deformed black-hole metrics,
subject to possible weak and strong quantum gravity effects, respectively, are
obtained, by a highly effective method. The accuracy and effectiveness of these expressions are then illustrated by 
numerically solving the differential equation governing the deflection angle directly in the weak quantum-gravity effect situation.

\medskip

\medskip

\medskip


\end{abstract}

\section{Introduction}
\setcounter{equation}{0}
\setcounter{figure}{0}
\setcounter{table}{0}

Gravitational  deflection of  light  is one of the three classical experimental tests of General Relativity proposed by Einstein himself. In
modern theoretical physics, these
tests and their computational realizations are of relevance and interest in the studies of various extended and modified theories as well, developed to 
enrich and improve Einstein's theory.
In these 
 extended situations, it is often difficult to determine the deflection or bending angle with full precision and suitable approximations are
inevitable.  Among these, explicit calculations \cite{BH,BIS,BBD,BBH,BS,CM,E,EA,FS,IH,IP,MT,S,T,ZX} may involve evaluating some complicated integrals and
implicit calculations \cite{ADGH,AGS,AHGH,AP,ALS,F,HHY,LC,Q,SB,XY} amount to solving some sophisticated nonlinear equations. 
Recently, in \cite{LC}, a study on the determination of the bending angle of light deflection subject to
deformed Schwarzschild metrics taking account for possible weak and strong
quantum gravity effects is conducted. Phenomenologically, in such a weak quantum gravity effect situation, a deviation is turned on by the presence
of a small parameter, $\kappa>0$, which serves to deform the usual Schwarzschild black hole metric. (Since it would be interesting to know whether the deformed metric might return any measurable effects, any information on an order-of-magnitude estimate for $\kappa$ would be useful. For such, we may refer to
\cite{CL} for a discussion within the framework of the Randers--Finsler asymmetric spacetime \cite{R,GLS}.) Specifically, along such a formalism, it is shown \cite{LC} that
the bending angle assumes the form
\be
\hat{\alpha}=2\left(2+\frac\kappa4\right)\frac{GM}\xi,\label{1.1}
\ee
where $G$ is Newton's gravitational constant, $M$ the mass of a radially symmetric gravitational source, $\xi$ the 
 the distance of the closest approach,
and the speed of light in vacuum is taken to be unity.
Furthermore, in a strong quantum gravity effect situation, a deviation of the gravitational metric is given in terms of a positive integer, $n$, so
that the bending angle is shown \cite{LC} to follow the formula
\be
\hat{\alpha}=4(1+4n)\frac{GM}\xi.\label{1.2}
\ee
Both formulas are seen to overwhelm the classical
 Einstein angle, $\theta_{\mbox{E}}=\frac{4GM}\xi$. (Part of the motivation of the study \cite{LC} is to account for some recently observed data 
which deviate away \cite{CRM} from those predicted classically based on the Einstein theory.)  In order to derive these formulas, in \cite{LC},  it is first shown, in
each case, that the radial variable and azimuthal angle satisfy
a Friedmann-type differential equation. In order to overcome this difficult structure, the equation is further differentiated and a linearization then taken. 
Solving the linearized equation leads to a relation between the radial variable and the azimuthal angle. Using the leading-order approximation of this relation
in the second-order differential equation obtained from differentiating the Friedmann-type equation and taking approximation again a nonlinear functional equation
is obtained. Finally solving the leading-order approximation of this last equation results in the bending angle. Thus, we have seen that, in order to find the bending angle, many steps of approximations are taken and the errors so accumulated are hard to keep track of. In fact, such an approach is well known and widely used in literature (cf.\cite{MTW,PK}). On the other hand, since in
the context of
gravitational scales, quantum-gravity effects are often small compared with the underlying classical ones, it will be useful and interesting to know 
detailed properties of the bending angle with regard to its higher-order terms.
Notably, in \cite{BW}, an analytic calculation of the bending angle in general relativity is carried out 
to the second order in $\frac{GM}\xi$. In the Schwarzschild coordinates, their result reads
\be\label{1.3}
\hat{\alpha}=\frac{4GM}\xi+\left(\frac{15\pi}4-4\right)\left(\frac{GM}\xi\right)^2,
\ee
and, in \cite{BBD,BBH}, based on a semiclassical calculation, it is found that the bending angle is given by
\be\label{1.4}
\hat{\alpha}=\frac{4GM}b+\frac{15\pi}4\left(\frac{GM}b\right)^2+c_b\hbar \,\frac{G^2 M}{b^3},
\ee
 where $\hbar$ is the Planck constant, $b$ the impact parameter, and $c_b$ a $b$-dependent
quantity, which coincides with \eqref{1.3} up to the second order in $\frac{GM}\xi$ in the classical gravity limit, $\hbar=0$,
in the Schwarzschild coordinates. This last formula is seen to take a clear quantum-gravity-model departure from its classical limit (\ref{1.3}). See \cite{BW,BBH} for detail
 and also \cite{FF,KP,FGG,AK} for some other studies on the fine structures of the bending angle. In view of these studies, it will be interesting to uncover the possibly hidden higher-order terms in the bending angle formulas  (\ref{1.1}) and (\ref{1.2}),
so that both classical and quantum gravity effects, as well as their interplay, are clearly exhibited, through the bending angle.
Indeed, in the current work, we set forth to extend the study in \cite{LC}, to get a full determination of the bending angle of light deflection, subject to 
the described weak and strong quantum effects, within controlled approximations.
In doing so we will be able to obtain precise information, in principle, regarding the fine structures of the bending angle, containing all second- or higher-order terms. In particular,
we see that the linear-approximation results (\ref{1.1}) and (\ref{1.2}) are actually underestimates of the bending angle in both situations, and hence, all second- or higher-order additional terms serve to contribute to getting  more accurate knowledge of the bending angle.
 Methodologically, comparing with that in \cite{LC,MTW,PK}, our approach
 is more direct and effective in that we
work directly on the integration of the Friedmann-type equations without taking further approximations. The integral in each case assumes a difficult form. However, we will show
that its appropriately centered Taylor expansion is quite manageable to allow well-controlled calculations, thus providing precise information in the Taylor expansion
and associated truncation errors. This method henceforth enables a determination of the bending angle of the problem with any desired accuracy threshold, within
our approximations.

We should note that the bending angle may be expressed in terms of the distance of the closest approach $\xi$ such as in \eqref{1.1}--\eqref{1.3} or in terms of the impact parameter $b$ such as  in \eqref{1.4}. Although the former is 
coordinate-dependent, it may be made an observable quantity such as that in the Schwarzschild coordinates. The latter on
the other hand is
coordinate-independent and thus of obvious advantage and meaningfulness in general relativity. In practice, both $\xi$ and $b$
are popularly used in calculating the bending angle and the results may be converetd into each other once
one establishes the relation between $\xi$ and $b$ as in \cite{MTW,BW,KP,Wald}. At textbook levels, both
$\xi$ and $b$ are used in \cite{MTW,Wald} but only $\xi$ is used in \cite{Weinberg,FN}, and, in the research literature cited here, 
$b$ is used dominantly in \cite{BBD,BBH,FF,KP,AK} and $\xi$ in \cite{LC,BW,FGG}. Since our work is directly related to that in
\cite{LC}, we shall use $\xi$ to represent various expressions obtained for the bending angle such that  explicit comparisons 
between our results and those in \cite{LC} are
readily accessible. 

The content of the rest of the paper is as follows. In Section 2 we calculate the bending angle, $\hat{\alpha}$, subject to possible weak quantum gravity effects. We first review the phenomenologically deformed Schwarzschild line element  following \cite{LC}
and arrive at a nonlinear equation governing the radial variable. The complexity of this equation does not allow an explicit
calculation of the bending angle and an approximation is necessary. The work of Section 2 is based on a  linear approximation of this
equation.
We then obtain full-structure formulas for $\hat{\alpha}$ based on a linear-in-$\kappa$ approximation as in \cite{LC} and on the full equation
without any further approximation. We shall see that, in doing so, the bending angle formula is improved and refined. 
In Section 3 we calculate $\hat{\alpha}$ subject to possible strong quantum gravity effects. We show that this situation allows a complete
determination of $\hat{\alpha}$ in the sense that all coefficients in the Taylor expansion of the integral that gives rise to $\hat{\alpha}$ may be computed
explicitly. As an example, we present an expression for $\hat{\alpha}$ with a 4th-order truncation error in $\frac{GM}\xi$ whose leading term is as stated in (\ref{1.2}).  In Section 4, we reconsider the weak quantum gravity situation  and solve the concerned nonlinear equation by a quadratic equation
approximation. In the weak quantum-gravity effect situation, we satisfactorily observe how our subsequently enhanced approximations steadily and
monotonically improve and refine the results. In Section 5, we carry out a numerical integration of the full differential equation
governing the deflection angle, aimed at demonstrating the accuracy and effectiveness of our defection angle formulas based on
the approximations of various orders, for the weak quantum-gravity situation, whose analytic structure is complicated in that the
differential equation is ``fully nonlinear". Fortunately,  the equation enables us to come up with an effective iterative algorithm to
compute the exact solution for the deflection angle so that we may compare the results obtained from approximations and 
solving the full equation and observe the anticipated monotone convergence.
 In all situations, our bending angle formulas contain (\ref{1.1}), (\ref{1.2}), and (\ref{1.3}) as limiting results.
 In the ending paragraph, we conclude the article with a summary. We note that, in order to facilitate our calculation, we have
benefited from and resorted to 
the symbolic computational tools provided by MAPLE 10.

\section{Light deflection subject to weak quantum gravity effects  based on linear approximation}
\setcounter{equation}{0}
\setcounter{figure}{0}
\setcounter{table}{0}

Following \cite{LC}, within the framework of the Finsler geometry \cite{R,GLS,Ant,Bao}, the line element of a phenomenological spacetime subject to weak quantum-gravity effect
characterized by a small dimensionless parameter $\kappa$, beyond the Schwarzschild sphere $r=2GM$,  is given by 
\be\label{2.1}
\dd s^2=a\dd t^2 -\frac1a\dd r^2-r^2(\sin^2\theta\,\dd\phi^2+ \dd\theta^2)+\kappa\sqrt{a}\frac{GM}r\left(1-a\frac{J^2}{E^2 r^2}\right)^{\frac34}\dd t
\sqrt{\dd t\dd r},
\ee
where $G$ is the Newton gravitational constant, $M$ the mass of a radially symmetric gravitational source, $a=1-\frac{2GM}r$ the Schwarzschild factor, $r$
the radial coordinate, $\theta$ the colatitude or polar angle coordinate, $\phi$ the longitude or azimuthal angle coordinate, $t$ time,
and $E,J$ are some constants, giving rise to the energy and angular momentum per unit mass of the particle  \cite{LC}.
As in \cite{LC}, the parameter $\kappa$ is a dimensionless quantity which is switched on to take account of the quantum-gravity
effect. On the other hand, however, it should be noted that the metric \eqref{2.1} is simply a deformed Schwarzschild metric and our study and discussion of light deflection subject to it in this paper is understood to be classical rather than quantum-mechanical.

 When the motion of the particle is assumed to be confined in the equatorial plane
$\theta=\frac\pi2$, the line element (\ref{2.1}) becomes
\be\label{2.2}
\dd s^2=a\dd t^2 -\frac1a\dd r^2-r^2\,\dd\phi^2+\kappa\sqrt{a}\frac{GM}r\left(1-a\frac{J^2}{E^2 r^2}\right)^{\frac34}\dd t
\sqrt{\dd t\dd r}.
\ee
To proceed further, we use $\tau$ to denote a generic trajectory coordinate variable and dot the corresponding derivative with respect to $\tau$. Then 
the null condition $\dd s^2=0$ for the light-like motion of the particle leads to \cite{LC}:
\be\label{2.3}
a{\dot{ t}}^2 -\frac{{\dot{ r}}^2}a-r^2 \,{\dot{\phi}}^2+\kappa\sqrt{a}\frac{GM}r\left(1-a\frac{J^2}{E^2 r^2}\right)^{\frac34}
{\dot{ t}^{\frac32}}{\dot{ r}^{\frac12}}=0.
\ee
On the other hand, integration of the autoparallel geodesic equations resulting from the line element (\ref{2.1}) leads to the conservation laws \cite{LC}:
\bea
a{\dot{t}}+\kappa\sqrt{a}\frac{3GM}{4r}\left(1-a\frac{J^2}{E^2 r^2}\right)^{\frac34}\sqrt{\dot{t}\dot{r}}&=&E,\label{2.4}\\
r^2\dot{\phi}&=&J.\label{2.5}
\eea
In view of (\ref{2.4}) and (\ref{2.5}), we see that the equation (\ref{2.3}) becomes
\be\label{2.6}
\dr^2=-\frac13 a^2\dt^2+\frac43 Ea\dt-\frac{aJ^2}{r^2}.
\ee
On the other hand, (\ref{2.4}) may be solved for $\dt$ to give us
\be\label{2.7}
\sqrt{\dt}=\frac12\sqrt{\frac9{16a}\left(\kappa\frac{GM}r\right)^2 \left(1-a\frac{J^2}{E^2 r^2}\right)^{\frac32}\dr+\frac{4E}a}-\frac3{8\sqrt{a}}\kappa\frac{GM}r
\left(1-a\frac{J^2}{E^2 r^2}\right)^{\frac34}\sqrt{\dr}.
\ee
Rewriting this relation as $\dt=f(\dr)$, we see that the equation (\ref{2.6}) assumes the form
\bea\label{2.8}
\dr^2&=&-\frac13 a^2 f^2(\dr)+\frac43 Ea f(\dr)-\frac{aJ^2}{r^2}\nn\\
&\equiv& g(\dr),
\eea
which is still too complicated to solve. Nevertheless, since $\eta=\kappa\frac{GM}r$ is small, we may expand the expression $\dr^2-g(\dr)$ around $\eta=0$ to
obtain
\be\label{2.9}
\dr^2-g(\dr)=\dr^2-E^2+\frac{aJ^2}{r^2}+\frac{\kappa E^{\frac32}}2\left(1-a\frac{J^2}{E^2 r^2}\right)^{\frac34}\frac{GM}r \sqrt{\dr}+\mbox{O}(\eta^3).
\ee 
Note that the quadratic term in such an expansion is absent such that the linear part already achieves a high-accuracy (second-order)
approximation. Note also that, in the classical gravity limit $\kappa=0$, (\ref{2.3})--(\ref{2.5}) lead to the solution
\be\label{2.10}
\dr_0\equiv \dr\,|_{\kappa=0}=\sqrt{E^2-\frac{aJ^2}{r^2}}.
\ee
Thus, expanding the right-hand side of (\ref{2.9}) around the classical solution (\ref{2.10}), we get
\be\label{2.11}
\dr^2-g(\dr)={\dr_0^{2}}\,\frac{\kappa GM}{2r}+\dr_0\left(2+\frac{\kappa GM}{4r}\right)(\dr-\dr_0)
+\mbox{O}([\dr-\dr_0]^2)+\mbox{O}(\eta^3).
\ee
Neglecting the higher-order error terms in (\ref{2.11}), we can solve the equation $\dr^2-g(\dr)=0$ to obtain the solution
\be\label{2.12}
\dr=\dr_0\left(1-\frac{\kappa GM}{8r}\right)\left(1+\frac{\kappa GM}{8r}\right)^{-1}.
\ee

Of course, one may take further approximations of (\ref{2.12}) in order to facilitate the computation. First, since $\eta=\kappa\frac{GM}r$ is small, 
we may use a linear truncation in (\ref{2.12}) in terms of $\eta$ which enables us to arrive at
\be\label{2.14}
\dr=\dr_0\left(1-\kappa\frac{GM}{4r}\right).
\ee
This equation is what  studied in \cite{LC}. We will focus on (\ref{2.14}) first as a computational illustration.

Following \cite{LC}, we insert (\ref{2.5}) into (\ref{2.14}) to arrive at
\be
\frac1{r^2}\frac{\dd r}{\dd \phi}=\sqrt{\frac{E^2}{J^2}-\frac a{r^2}}\left(1-\kappa\frac{GM}{4r}\right).
\ee
Thus, with $u=\frac{GM}r$, the above equation conveniently becomes \cite{LC}:
\be\label{2.16}
\frac{\dd u}{\dd\phi}=-\sqrt{\left(\frac{EGM}J\right)^2-u^2(1-2u)}\left(1-\frac{\kappa u}4\right).
\ee
Since the light ray is assumed to pass around the gravitational source at the shortest distance $r=\xi$ (the  
 the distance of the closest approach), where $\frac{\dd r}{\dd\phi}=0$ or
$\frac{\dd u}{\dd\phi}=0$ and $u_0=\frac{GM}{\xi}$, we
obtain from (\ref{2.16}) the result
\be\label{2.17}
\left(\frac{EGM}J\right)^2=u_0^2(1-2u_0),
\ee
which fixes the ratio of $E$ and $J$ as a by-product. Substituting (\ref{2.17}) into (\ref{2.16}), integrating (\ref{2.16}), and noting  the correspondence
\be\label{2.18}
u=0,\quad \phi=\pm\left(\frac\pi2+\alpha\right);\quad u=u_0,\quad \phi=0,
\ee
between the variable $u$ and the azimuthal angle $\phi$ where $\hat{\alpha}=2\alpha$ is the angle of light ray deflection, we see that 
the branch $0<\phi<\frac{\pi}2$ is given by the integral
\be\label{2.19}
\phi(u)=-\int_{u_0}^u\frac{\dd u'}{\sqrt{u_0^2(1-2u_0)-{u'}^2(1-2u')}\left(1-\frac{\kappa u'}4\right)},\quad 0<u<u_0.
\ee
To proceed further, we set 
\be\label{2.20}
p(u)=\left(u^2_0(1-2u_0)-u^2(1-2u)\right)\left(1-\frac{\kappa u}4\right)^2,
\ee
and $u=u_0 v$. Then we have
\bea\label{2.21}
\frac\pi2+\alpha&=&\int^{u_0}_0\frac{\dd u}{\sqrt{p(u)}}\nn\\
&=&\int_0^1\frac{\dd v}{\sqrt{q(u_0,v)}}\equiv Q(u_0),
\eea
where
\be
q(u_0,v)=\left(1-2u_0-v^2(1-2u_0 v)\right)\left(1-\frac{\kappa u_0v}4 \right)^2.
\ee
It remains to compute $Q(u_0)$ effectively for $u_0>0$. Of course, we have
\be\label{2.23}
Q(0)=\int_0^1\frac{\dd v}{\sqrt{1-v^2}}=\frac\pi2.
\ee
Besides, we also have
\bea
Q'(0)&=&\int_0^1\left(\frac{1-v^3}{({1-v^2})^{\frac32}}+\frac{\kappa v}{4\sqrt{1-v^2}}\right)\,\dd v=2+\frac{\kappa}4,\label{2.24}\\
Q''(0)&=&\int_0^1 \left(\frac{3(1-v^3)^2}{(1-v^2)^{\frac52}}+\frac\kappa2\frac{(1-v^3)v}{(1-v^2)^{\frac32}}+\frac{\kappa^2}8\frac{v^2}{\sqrt{1-v^2}}\right)\,\dd v\nn\\
&=&\left(\frac{15\pi}4-4\right)+\frac12\left(\frac{3\pi}4-1\right)\kappa+\frac{\pi}{32}\kappa^2.\label{2.25}
\eea
In principle, there is no difficulty in getting the values of derivatives of $Q$ at $u_0=0$ of any orders such that the exact value of $Q(u_0)$ may be estimated
within arbitrary accuracy. It is interesting that all such values stay positive for any $\kappa\geq0$ so that we always approximate the true value of $Q(u_0)$ from below. In fact,
as an illustration, we similarly obtain
\be\label{2.26}
Q'''(0)=\left(122-\frac{45\pi}2\right)+\frac94\left(5-\pi\right)\kappa+\left(1-\frac{3\pi}{16}\right)\kappa^2+\frac1{16}\kappa^3,
\ee
which stays positive for all $\kappa\geq0$. Since $u_0$ is small, we are ensured with $Q'''(v)>0$ for any $v\in (0,u_0)$. Thus, in view of (\ref{2.21}), (\ref{2.23})--(\ref{2.26}), we get the following formula for the deflection or bending angle:
\bea\label{2.27}
\hat{\alpha}&=&2Q'(0) u_0+Q''(0) u_0^2+\frac13 Q'''(v) u_0^3\quad\quad (\mbox{some } v\in(0,u_0))\nn\\
&=&2\left(2+\frac\kappa4\right)\frac{GM}\xi+\left(\left[\frac{15\pi}4-4\right]+\frac12\left[\frac{3\pi}4-1\right]\kappa\right)\left(\frac{GM}\xi\right)^2+\mbox{O}(\kappa^2)+\mbox{O}\left(\left[\frac{GM}\xi\right]^3\right),\quad \quad 
\eea
in which the linear part, in $\frac{GM}\xi$, is as stated in (\ref{1.1}), obtained in \cite{LC}, the quadratic correction is new, 
whose leading term is as given in (\ref{1.3}), obtained in \cite{BW}, and the quadratic-in-$\kappa$ and cubic-in-$\frac{GM}\xi$ error terms are positive. Here, for consistency with the linear-in-$\kappa$ 
approximation (\ref{2.14}), we have omitted higher-order-in-$\kappa$ terms in the formula. Thus, even with the addition of
a positive quadratic correction, the estimate for the deflection angle is still an underestimate.

We next consider the sharper, full, equation (\ref{2.12}) without the linear truncation. With the same reformulation, we see that (\ref{2.16}) is now replaced with
\be\label{2.28}
\frac{\dd u}{\dd\phi}=-\sqrt{\left(\frac{EGM}J\right)^2-u^2(1-2u)}\left(1-\frac{\kappa u}8\right)\left(1+\frac{\kappa u}8\right)^{-1}.
\ee
Thus, as before, we arrive similarly at the formula
\be\label{2.29}
\frac\pi2+\alpha=\int_0^1\frac{\dd v}{\sqrt{q_1(u_0,v)}}\equiv Q_1(u_0),
\ee
where now
\be
q_1(u_0,v)=(1-2u_0-v^2(1-2u_0 v))\left(1-\frac{\kappa u_0 v}8\right)^2\left(1+\frac{\kappa u_0 v}8\right)^{-2}.
\ee
A direct computation gives us the results
\bea\label{2.31}
Q_1(0)&=&\frac\pi2,\quad Q_1' (0)=2+\frac\kappa4,\nn\\
 Q_1''(0)&=&\left(\frac{15\pi}4-4\right)+\frac12\left(\frac{3\pi}4-1\right)\kappa+\frac\pi{64}\kappa^2,\nn\\
Q_1'''(0)&=&\left(122-\frac{45\pi}2\right)+\frac94\left(5-\pi\right)\kappa+\frac12\left(1-\frac{3\pi}{16}\right)\kappa^2+\frac1{64}\kappa^3,
\eea
All these quantities are again positive. Thus, as in (\ref{2.27}), we have
\be\label{2.32}
\hat{\alpha}=2\left(2+\frac\kappa4\right)\frac{GM}\xi
+\left(\left[\frac{15\pi}4-4\right]+\frac12\left[\frac{3\pi}4-1\right]\kappa+\frac{\pi}{64}\kappa^2\right)\left(\frac{GM}\xi\right)^2
+\mbox{O}\left(\left[\frac{GM}\xi\right]^3\right),
\ee
which agrees with (\ref{2.27}) completely, plus a refined positive second-order-in-$\kappa$ term, and again provides an effective underestimate for the deflection angle. 
Note that there is no difficulty in finding all $Q_1^{(m)}(0)$ explicitly such that we may obtain all higher-order terms in $\hat{\alpha}$ as
illustrated in the above manner with well-described truncation errors.

{\em Note.} One may raise the question whether it would be fully consistent and effective already to work on the second-order approximation to the equation
(\ref{2.12}) instead since the truncation we take in (\ref{2.9}) in terms of $\eta$ is also quadratic. Indeed, we have examined such
an approximation where (\ref{2.12}) is replaced by the quadratic equation
\be\label{x2.32}
\dr=\dr_0\left(1-\frac14\frac{\kappa G M}{r}+\frac1{32}\left[\frac{\kappa GM}r\right]^2\right).
\ee
From (\ref{x2.32}) and following the same procedure, we see that the bending angle is now given by
\be
\frac\pi2+\alpha=\int_0^1\frac{\dd v}{\sqrt{q_2(u_0,v)}}\equiv Q_2(u_0),
\ee
where
\be
q_2(u_0,v)=(1-2u_0-v^2(1-2u_0 v))\left(1-\frac{\kappa u_0 v}4+\frac{\kappa^2 u_0^2 v^2}{32}\right).
\ee
Interestingly, indeed, $Q_2^{(m)}(0)$ agrees with $Q_1^{(m)}(0)$ (as listed in (\ref{2.31})) for $m=0,1,2$ but differs at $m=3$. In fact, $Q_2'''(0)$ contains
all the terms of $Q_1'''(0)$ listed in (\ref{2.31}) except the tail term of the order $\kappa^3$ is absent. This examination confirms
the expectation that, up to second-order terms, the second-order approximation of (\ref{2.12}),
namely (\ref{x2.32}), for the calculation of the bending angle, is as effective as the full equation (\ref{2.12}).

\section{Light deflection subject to strong quantum gravity effects}
\setcounter{equation}{0}

Following the phenomenological approach in \cite{LC}, we  consider in the equatorial plane the line element
\be
\dd s^2=a\dd t^2-\frac{(4n+1)^2}a\dd r^2-r^2\dd\phi^2,
\ee
where the integer $n=0,1,2,\dots$ is a  deformation parameter and $a$ the Schwarzschild factor defined in the previous section. 
When $n>0$, this metric gives rise to a positive energy density of the universe which falls off following an inverse-square law of the radial variable \cite{LC}.
Thus, as before, the null trajectory condition gives us the equation
\be\label{3.2}
a\dt^2-\frac{(4n+1)^2}a\dr^2-r^2\dot{\phi}^2=0.
\ee
On the other hand, it follows from integrating the autoparallel geodesic equations under the given line element that there are two additional 
conserved relations \cite{LC}:
\bea
a\dt&=&E,\label{3.3}\\
r^2\dot{\phi}&=&J,\label{3.4}
\eea
with $E,J$ two positive parameters. Thus, inserting (\ref{3.3}) and (\ref{3.4}) into (\ref{3.2}), we
arrive at the exact equation \cite{LC}:
\be\label{3.5}
(4n+1)^2\dr^2=E^2-a\frac{J^2}{r^2}.
\ee
Therefore, with the same change of variable, $u=\frac{GM}r$,  the updated asymptotic correspondence
\be\label{3.6}
u=0,\quad \phi=\pm\left((4n+1)\frac\pi2+\alpha\right);\quad u=u_0,\quad \phi=0,
\ee
and the Friedmann-type differential equation
\be\label{3.7}
(4n+1)\frac{\dd u}{\dd\phi}=-\sqrt{\left(\frac{EGM}J\right)^2 -u^2(1-2u)},
\ee
we have as in (\ref{2.21}) the conclusion
\be\label{3.8}
(4n+1)\frac\pi2+\alpha=(4n+1)\int_0^1\frac{\dd v}{\sqrt{q_3(u_0,v)}}\equiv (4n+1)Q_3(u_0),
\ee
where we have used the relation (\ref{2.17}) to get
\be\label{3.9}
q_3(u_0,v)=1-2u_0-v^2(1-2u_0 v),
\ee
which is much simpler than those considered in the weak quantum-gravity-effect cases in the previous section. As a consequence, there is no difficulty to obtain
all $Q_3^{(m)}(0)$ explicitly among which
\bea\label{3.10}
Q_3(0)&=&\frac\pi2,\quad Q_3'(0)=2,\quad Q_3''(0)=\frac{15\pi}4-4,\nn\\
Q_3'''(0)&=&122-\frac{45\pi}2,\quad Q_3^{(4)}(0)=\frac{10395\pi}{16}-1560,
\eea
which are all positive. Hence the associated angle of deflection $\hat{\alpha}=2\alpha$ is given up to the third order (say) of $\frac{GM}\xi$ by
\be\label{3.11}
\frac{\hat{\alpha}}{4n+1}=4\left(\frac{GM}\xi\right)+\left(\frac{15\pi}4-4\right)\left(\frac{GM}\xi\right)^2+\left(\frac{122}3-\frac{15\pi}2\right)\left(\frac{GM}\xi\right)^3
+\mbox{O}\left(\left[\frac{GM}\xi\right]^4\right).
\ee
The first term on the right-hand side of (\ref{3.11}) is obtained earlier in \cite{LC} as stated in (\ref{1.2}). It is interesting that the second-order term on the right-hand side of (\ref{3.11}) is again as that given in (\ref{1.3}),  obtained in \cite{BW}. In fact, this term appears
in all the bending angle formulas we present in our current study.

\section{Bending angle subject to weak quantum gravity effects based on quadratic approximation}
\setcounter{equation}{0}

In Section 2, we have seen that the use of the full equation (\ref{2.12}) or the second-order-in-$\kappa$ approximation (\ref{2.32}) of the linear-in-$(\dr-\dr_0)$ approximation of the equation (\ref{2.9}), namely, (\ref{2.11}), gives us the formula
(\ref{2.32}) which is further improved and refined from that based on the solution of its linear-in-$\kappa$ approximate equation
 (\ref{2.14}) (which is as used in \cite{LC}), that is, (\ref{2.27}). Thus, it will be interesting to know what happens when we modify (\ref{2.12}) with
a further, yet,  improved approximation such as a quadratic-in-$(\dr-\dr_0)$ one. In this section, we investigate this issue.

First we note that the quadratic approximation of the right-hand side of (\ref{2.9}) in terms of $\dr-\dr_0$ is
\be\label{4.1}
\dr^2-g(\dr)={\dr_0^2}\,\frac{\kappa GM}{2r}+\dr_0\left(2+\frac{\kappa GM}{4r}\right)(\dr-\dr_0)
+\left(1-\frac{\kappa GM}{16r}\right)(\dr-\dr_0)^2+\mbox{O}([\dr-\dr_0]^3)+\mbox{O}(\eta^3).
\ee
Neglecting the truncation errors and solving the quadratic equation, we obtain
\be\label{4.2}
\dr=\dr_0\left(1-\frac2{16-\eta}\left[8-\sqrt{64-16\eta+3\eta^2}+\eta\right]\right),
\ee
where again
$
\eta=\frac{\kappa GM}r.
$
Thus, as in (\ref{2.21}), we have
\be\label{4.5}
\frac\pi2+\alpha=\int_0^1\frac{\dd v}{\sqrt{q_4(u_0,v)}}\equiv Q_4(u_0),
\ee
where
\be\label{4.6}
q_4(u_0,v)=(1-2u_0-v^2(1-2u_0 v))\left(1-\frac2{16-\eta}\left[8-\sqrt{64-16\eta+3\eta^2}+\eta\right]\right)^2,
\ee
where $\eta=\kappa u_0 v$. Interestingly, although (\ref{4.6}) appears complicated, there is no difficulty in obtaining $Q_4^{(m)}(0)$ for $m=0,1,2,3,\dots$,
among which we present
\bea\label{4.8}
Q_4(0)&=&\frac\pi2,\quad Q'_3(0)=2+\frac\kappa4,\nn\\
 Q''_4(0)&=&\left(\frac{15\pi}4-4\right)+\left(\frac{3\pi}8-\frac12\right)\kappa
+\frac\pi{32}\kappa^2,\nn\\
Q_4'''(0)&=&\left(122-\frac{45\pi}2\right)+\frac94(5-\pi)\kappa+\left(1-\frac{3\pi}{16}\right)\kappa^2+\frac7{128}\kappa^3.
\eea
Thus we are led to arrive at the formula
\be\label{4.10}
\hat{\alpha}=2\left(2+\frac\kappa4\right)\frac{GM}\xi
+\left(\left[\frac{15\pi}4-4\right]+\frac12\left[\frac{3\pi}4-1\right]\kappa+\frac{\pi}{32}\kappa^2\right)\left(\frac{GM}\xi\right)^2
+\mbox{O}\left(\left[\frac{GM}\xi\right]^3\right),\quad
\ee
for the bending angle, where the third-order error term is again positive, which indeed improves upon (\ref{2.32}), since all error terms there are also positive.
\medskip

Thus, in the weak quantum gravity effect situation, we have seen from our study in Sections 2 and 4 that, up to  second-order terms in $\frac{GM}\xi$, with our enhanced approximation approaches,  the results
for the bending angle are steadily and monotonically improved, as displayed in the subsequently obtained formulas (\ref{2.27}), (\ref{2.32}), and (\ref{4.10}).
 All these formulas contain
 (\ref{1.3}) as their classical gravity limit, when $\kappa=0$.
 
\section{Computation of the bending angle without approximation}
\setcounter{equation}{0}

In this section, we compute the bending angle in the weak quantum-gravity situation directly without resorting to approximation or error truncation.

First, note that, at the shortest approaching distance $\xi$ from the gravitational source where $u_0=\frac{GM}\xi$, we have by using $\frac{\dd r}{\dd\phi}=0$,
so that $\dr=0$, in (\ref{2.7}) that
\be\label{5.1}
f(\dr)=\dot{t}=\frac{E}{1-2u_0}.
\ee
Inserting (\ref{5.1}) into (\ref{2.8}), we arrive at (\ref{2.17}) again. This indicates that (\ref{2.17}) is valid in general.

Next, with $u=\frac{GM}r$ and $a=1-2u$, we see that (\ref{2.7}) becomes
\bea\label{5.2}
\sqrt{\dot{t}}=\sqrt{f}&=&\frac12\sqrt{\frac9{16(1-2u)}(\kappa u)^2\left(1-\frac{(1-2u) u^2 J^2}{(EGM)^2}\right)^{\frac32}\left(-\frac J{GM}\frac{\dd u}{\dd\phi}\right)+
\frac{4E}{1-2u}}\nn\\
&&-\frac3{8\sqrt{1-2u}}\kappa u\left(1-\frac{(1-2u)u^2 J^2}{(EGM)^2}\right)^{\frac34}\sqrt{-\frac J{GM}\frac{\dd u}{\dd \phi}}.
\eea
Furthermore, from (\ref{2.5}), we have
\be\label{5.3}
\frac{\dd r}{\dd\phi}=\frac{\dr}{\dot{\phi}}=\frac{r^2}J \dr.
\ee
Hence,  using (\ref{5.3}) in (\ref{2.8}), we have
\be\label{5.4}
\frac{\dd u}{\dd\phi}=-\frac{GM}J\sqrt{-\frac13 (1-2u)^2 f^2 +\frac43 E(1-2u)f-\frac{(1-2u)u^2 J^2}{(GM)^2}},
\ee
where $f$ is as given in (\ref{5.2}). 

In order to facilitate our computation, we normalize the variable by setting $u=u_0 v$. Then, in view of (\ref{2.17}), we can rewrite (\ref{5.4}) as
\be\label{5.5}
\frac{\dd v}{\dd\phi}=-\sqrt{1-2u_0}\sqrt{-\frac13(1-2u_0 v)^2 H^2+\frac43(1-2u_0 v)H-\frac{v^2 (1-2u_0 v)}{1-2u_0}},
\ee
where $H=\frac fE$ is defined by the corresponding formula
\bea\label{5.6}
\sqrt{H}&=&\frac12\sqrt{\frac{9(\kappa u_0 v)^2}{16(1-2u_0v)}\left(1-\frac{v^2(1-2u_0 v)}{1-2u_0}\right)^{\frac32}\left(-\frac1{\sqrt{1-2u_0}}\frac{\dd v}{\dd\phi}\right)+\frac4{1-2u_0 v}}\nn\\
&&-\frac{3\kappa u_0 v}{8\sqrt{1-2u_0 v}}\left(1-\frac{v^2 (1-2u_0 v)}{1-2u_0}\right)^{\frac34}\sqrt{-\frac1{\sqrt{1-2u_0}}\frac{\dd v}{\dd\phi}}.
\eea

Finally, with the afore-going preparation, we arrive at the differential equation
\bea
\frac{\dd v}{\dd\phi}&=&F\left(v,\frac{\dd v}{\dd \phi}\right),\quad \phi>0,\label{5.7}\\
v&=&1\quad\mbox{when }\phi=0,\label{5.8}
\eea
where the right-hand side of (\ref{5.7}) is defined by (\ref{5.5})--(\ref{5.6}) and the initial condition (\ref{5.8}) comes from $u=u_0$ at $\phi=0$. Thus, we are to 
integrate (\ref{5.7})--(\ref{5.8}) and find 
\be\label{5.9}
\phi_0=\frac\pi2+\alpha
\ee
for some $\alpha>0$ where $v$ vanishes. Then $\hat{\alpha}=2\alpha$ is the bending angle. This initial-value problem appears rather complicated because (\ref{5.7}) is {\em fully nonlinear} in the sense that the equation is nonlinear in its (highest-order) derivative. However, it has
the nice features that (i) it is autonomous, and (ii) it does not contain free parameters but only the quantum deformation constant $\kappa$ and the classical parameter
$u_0=\frac{GM}\xi$.

To proceed within the above formalism, we note that $\frac{\dd v}{\dd\phi}=0$ at $\phi=0$. Thus, with this observation, we are led to the following
finite-difference iterative scheme with the uniform step $h=\phi_{k}-\phi_{k-1}$ for $k=1,2,\dots$ and the discretized values of $v$ at the corresponding $\phi_k$
given by
\bea
v_k'&=&F(v_k,v'_k),\quad k=0,1,2,\dots,\label{5.10}\\
v_0&=&1,\label{5.11}\\
\frac{v_{k+1}-v_{k}}h&=&v_k',\quad k=0,1,2,\dots.\label{5.12}
\eea
Note that $v_k'$ in (\ref{5.10}) is to be implicitly determined. Since $\kappa>0$ is small, we may obtain $v_k'$ at each step $k=0,1,2,\dots$ iteratively by the scheme
\be\label{5.13}
w_{n+1}=F(v_k,w_n),\quad n=0,1,2,\dots,
\ee
so that $v_k'$ is chosen to be the unique fixed point $w$ of the function $F(v_k,\cdot)$, or $w=F(v_k,w)$.
Here, in (\ref{5.13}), we may take $w_0=0$ or $w_0=F(v_k,0)$ as initial state at each step $k$. After $v_k'$ is obtained,  we get $v_{k+1}$ by (\ref{5.12}). Then we
repeat the same computation at the next step, $k+1$, as described.

To implement the computation, we note that $F(1,0)=0$ in (\ref{5.7}). That is, $v\equiv1$ is an equilibrium of the equation, which adds difficulty to start our iterative
scheme. However, we also note that, setting $\frac{\dd v}{\dd \phi}=0$ on the right-hand of (\ref{5.7}), we arrive at its approximate equation
\be\label{5.14}
\frac{\dd v}{\dd \phi}=-\sqrt{(1-2u_0)-v^2(1-2u_0 v)},\quad v(0)=1,
\ee
which still defies an explicit  integration. To facilitate our process,  we consider a further approximation of (\ref{5.14}) as follows:
\be
\frac{\dd v}{\dd\phi}=-\sqrt{2(1-3u_0)(1-v)},\quad v(0)=1,
\ee
which comes from taking the first-order approximation around $v=1$ of the cubic function of $v$ under the square root on the right-hand side of the
equation
whose solution is
\be
v=1-\frac{(1-3u_0)}2\phi^2.
\ee
Thus, in our actually implementation of the iterative scheme, we may choose
\be\label{5.17}
v_1=1-\frac{(1-3u_0)}2\phi^2_1,\quad v_1'=-(1-3u_0)\phi_1,
\ee 
as the initial state, at $k=1$ in the scheme (\ref{5.10})--(\ref{5.12}),  instead of starting at $k=0$. Another practical advantage of such an approach is that this enbles us to effectively replace the implicit scheme (\ref{5.10})--(\ref{5.12}) by the explicit (iterative)
one:
\be\label{5.18}
v'_{k+1}=F(v_k, v'_k),\quad k=1,2,\dots,
\ee
where $v_1$ and $v_1'$ are given in (\ref{5.17}). In our concrete implementation of numerics, both
the implicit method (\ref{5.10})--(\ref{5.12}) and the explicit (\ref{5.18}) have been tested, and, the latter is much less time
consuming and yields equally satisfactory results. Moreover, since (\ref{5.17}) actually enables a two-step iterative algorithm 
to our disposal, we have also conducted several relevant computational tests and found the scheme given by
\be\label{5.19}
v_{k+1}=v_{k-1}+2h F\left(v_k,\frac{v_k-v_{k-1}}h\right),\quad k=1,2,\dots,
\ee
which is iteratively of a central-difference type,
 to give rise to faster convergence. Thus, in our subsequent discussion, we will  present our results based on
the scheme (\ref{5.19}).

We will compute some examples based on the data of Sun with
\be\label{5.20}
M=1.989\times 10^{30} \,\,\mbox{(kg)}, \quad \xi=6.963\times 10^8\,\,\mbox{(m)}.
\ee
Using $G=6.674\times10^{-11}$ ($\mbox{m}^3\mbox{kg}^{-1}\mbox{s}^{-2}$) and $c=2.998\times10^8$ ($\mbox{ms}^{-1}$)
so that the line element is to be rescaled at suitable places resulting in the updated Schwarzschild radius $r_{\mbox{s}}=\frac{2GM}{c^2}$, etc.
With these, the 
Einstein deflection angle assumes the value
\be
\hat{\alpha}_{\mbox{E}}=
\frac{4GM}{c^2\xi}=8.484403697763715\times10^{-6} \,\,\mbox{(radians)}=1.750033884841381
 \,\,\mbox{(arcseconds)}.
\ee
Our results,  approximate and computed, should all exceed this value for arbitrary $\kappa>0$. On the other hand, to be
consistent with the studies \cite{BB,KK,Ki,Ali,Cal,Bar}, the parameter $\kappa$ should be a diminutive quantity in the order of the Planck length $\ell_{\mbox{P}}$ squared, where 
\be
\ell_{\mbox{P}}=\sqrt{\frac {\hbar G}{c^3}}\approx 1.616229\times10^{-35} \,\,\mbox{(m)}.
\ee
Thus, the approximate formulas (\ref{2.27}), (\ref{2.32}), and (\ref{4.10}) for the deflection angle $\hat{\alpha}$ are quite 
accurate and in fact indifferentiable for the data of Sun given in (\ref{5.20}). Nevertheless, it will be valuable and interesting to 
compare these approximate results with the exact results computed based on solving the initial-value problem of the full differential equation, consisting
of (\ref{5.7}) and (\ref{5.8}), with the finite-difference methods described earlier. In Table \ref{T}, we present a series of results
for $\kappa=0.001, 0.01, 0.05,0.08,0.1$, respectively. These results are obtained with a uniform step size $h=1.06\times 10^{-8}$ and a uniform computational termination threshold set at $8\times 10^{-9}$, balancing considerations on convergence and speed.
We see clearly that as we go from the left to right the results are gradually
and monotonically
improved with the right-most results the exact ones without truncation approximation and that as we move from the  top to bottom
the results increase monotonically as $\kappa$ increases. These resuts are exactly what expected mathematically as we have discussed.

\begin{table}
\centering

\caption{The deflection angle $\hat{\alpha}$ in arcseconds computed for Sun based on various approximation formulas and
direct differential equation computation without approximation with respect to different choices of the
metric modification parameter $\kappa$.}\label{T}
\medskip

\begin{tabular}{cccccr|r|r|r|r|r|r|r|r|}
 \hline
 $ \kappa $ & $\hat{\alpha}$ given by (\ref{2.27})& $\hat{\alpha}$ given by (\ref{2.32}) & $\hat{\alpha}$ given by (\ref{4.10})&  $\hat{\alpha}$ computed with (\ref{5.19})
\\
 \hline
 
  0.001    &1.750259860445759    &1.750259860445804    &1.750259860445850   &1.750365251152817 \\

  0.01    &1.752228654229671     &1.752228654234227    &1.752228654238782   &1.754738065040664 \\

 0.05    &1.760978848824838    &1.760978848938721     &1.760978849052603   &1.763483692816357 \\

 0.08    &1.767541494771213    &1.767541495062752     &1.767541495354292   &1.767856506704204 \\

 0.1    &1.771916592068796     &1.771916592524327     &1.771916592979857   &1.772229320592051 \\

\hline
\end{tabular}
\end{table}

We now compare the approximate results  (\ref{2.27}), (\ref{2.32}), and (\ref{4.10}) at $\kappa=0$ (in classical gravity situation), say $\hat{\alpha}_{\mbox{A}}$, with the classical linear result of Einstein
$\hat{\alpha}_{\mbox{E}}$. In our notation, we have
\be\label{5.23}
 \hat{\alpha}_{\mbox{E}}=4u_0,\quad\hat{\alpha}_{\mbox{A}}=4u_0+\left(\frac{15\pi}4-4\right)u^2_0.
\ee
 In Table \ref{T2}, we list some results in the interval
\be
\frac{GM}{c^2 \xi}\leq u_0\leq \frac{150GM}{c^2 \xi}\quad\mbox{(say)},
\ee
where $M$ and $\xi$ are as given in (\ref{5.20}). The reason for doing so is that usually a white dwarf may assume a mass comparable to that of Sun
and volume comparable to that  of Earth, e.g., Sirius B, whose radii are 0.76 to 0.86 in hundredths of those of Sun, with a mass
of 1.05 times of that of Sun, according to a report of Thejll and Shipman \cite{Sirius}. We may write down the associated relative error 
up to five decimal places as follows:
\bea\label{5.25}
E(u_0)&=&\frac{\hat{\alpha}_{\mbox{A}}-\hat{\alpha}_{\mbox{E}}}{\hat{\alpha_{\mbox{E}}}}=\left(\frac{15\pi}{16}-1\right)u_0\nn\\
&=&4.12621\times10^{-6},\quad
2.47556\times10^{-4},\quad 4.12606\times10^{-4},\quad 6.18909\times10^{-4},
\eea
for
\be\label{5.26}
u_0=\frac{GM}{c^2\xi},\quad\frac{60GM}{c^2\xi},\quad\frac{100GM}{c^2\xi},\quad\frac{150GM}{c^2\xi},
\ee
respectively. In \cite{FKLB}, it is reported that by using the Very Long Baseline Array (VLBA) a high accuracy  measurement
of the deflection angle by Sun may be achieved with a relative error bound below $3\times 10^{-4}$. Thus it is clear that the variations
due to quadratic corrections of the first two
predicted deflection angles listed in (\ref{5.25})--(\ref{5.26}) may not be detectable based on the VLBA method but those of the last two should be.

\begin{table}
\caption{The deflection angle $\hat{\alpha}$ in arcseconds computed for a
star of the solar mass  and a radius of one in hundred-fiftieth of the solar radius, based on the linear Einstein approximation and
quadratic approximation given in (\ref{5.23}), aimed at examining detectability of  quadratic corrections. }\label{T2}
\medskip

\centering
\begin{tabular}{cccccr|r|r|r|r|r|r|r|r|r|}
 \hline
 $u_0$ in the multiple of $\frac{GM}{c^2\xi}$ in (\ref{5.23})         &$\hat{\alpha}_{E}$  &  &$\hat{\alpha}_{A}$  
\\
 \hline
  $1$      &1.750033884841381&   &1.750041105580880\\

  $10$    &17.50033884841381&    &17.50106092236370\\

  $30$    &52.50101654524143&    &52.50751521079052\\

   $60$   &105.0020330904829&   &105.0280277526792\\

   $100$   &175.0033884841381&   &175.0755958791281\\
    
   $150$   & 262.5050827262071&  &262.6675493649346\\

\hline
\end{tabular}
\end{table}

\medskip
\medskip

In summary, we have presented a series of exact results which determine with high accuracy the bending angle of light deflection arising from two phenomenologically
proposed deformed black-hole metrics taking account for weak and strong
quantum gravity effects. Our method is direct and effective and provides precise information about the detailed properties of the bending angle and its
truncation errors so that the underlying effects including their interplay are clearly exhibited through these exact
formulas, which may be obtained based on a series of subsequently improved approximation approaches in the weak quantum-gravity effect situation, and is exact
in the strong situation, for the bending angle. In the more complicated weak quantum-gravity effect situation, we have also computed the bending angle by numerically solving the governing fully nonlinear differential equation based on an explicit two-step iterative algorithm and
demonstrated the accuracy and effectiveness of the bending angle formulas obtained with various orders of truncation errors. 
\medskip
\medskip

{\small{The authors would like to thank two anonymous referees whose comments and suggestions helped improve the presentation of the
paper. 
Xu's research was partially supported by National Natural Science Foundation of China under Grant No. 12071111 and
Yang's 
by National Natural Science Foundation of China under Grant No. 11471100. }}


\end{document}